\documentstyle[iopconf1]{article}
\def\ltap{\raisebox{-.4ex}{\rlap{$\sim$}} \raisebox{.4ex}{$<$}}
\def\gtap{\raisebox{-.4ex}{\rlap{$\sim$}} \raisebox{.4ex}{$>$}}
\newcommand{\Rsl}{{\not \! \! {R}}}
\newcommand{\Rslash}{{\not \! {R}}}
\begin{document}

\begin{flushright}
SINP/TNP/99-30 \\\texttt{hep-ph/9909566} 
\end{flushright}

\title{Direct $CP$ violation in $B$ decays in $R$-parity violating
models }

\author{Gautam Bhattacharyya
\footnote{E-mail: gb@tnp.saha.ernet.in ~~[{\it Invited Talk
presented at ``Beyond the Desert 1999'', Castle Ringberg, Tegernsee,
Germany, 6-12 June 1999.}]}
}

\affil{
Saha Institute of Nuclear Physics, 1/AF Bidhan Nagar, \\Calcutta 
700064, India}

\beginabstract 
In the standard model, $CP$ asymmetries in the $B^\pm \rightarrow
\pi^\pm K$ channels are $\sim 2\%$ based on perturbative
calculation. Rescattering effects might enhance it to at most $\sim
(20-25)\%$. We show that lepton-number-violating $\lambda'$ couplings
in supersymmetric models are capable of enhancing it to as large as
${\cal{O}}(100\%)$. Upcoming $B$ factories will test this scenario.

\endabstract

\centerline{
{\small Based on a work done in collaboration with A. Datta
\cite{bd}.}}

\section{Introduction to direct $CP$ violation}

Measurements of $CP$ violation in the upcoming $B$ factories could
reveal new physics with new phases. The best places to look for those
are some $CP$ asymmetries which in the standard model (SM) are
predicted to be too small, but nonetheless are going to be measured
with high precision. Measurements significantly larger than the SM
predictions will definitely point towards new physics with new
phases. The decay $B^+ \rightarrow \pi^+ K^0$ (at the quark level
$\bar{b} \rightarrow \bar{s} d\bar{d}$) constitutes one such mode. To
develop the formalism, let us consider a generic decay process
$B^+\rightarrow f$. The amplitude can be written as $A(B^+\rightarrow
f) = \sum_i|A_i|e^{i\phi_i^{W}}e^{i\phi_i^{S}}$. The summation implies
that in general there could be more than one diagram, labelled by the
index $i$, contributing to this process, and $\phi_i^{W}$ and
$\phi_i^{S}$ are the weak and strong phases, respectively, for the
$i$th diagram. The $CP$ asymmetry is defined as $a_{CP} \equiv
[{\cal{B}}(B^+\rightarrow f) - {\cal{B}}(B^-\rightarrow
\bar{f})]/[{\cal{B}}(B^+\rightarrow f) + {\cal{B}}(B^-\rightarrow
\bar{f})]$. Requiring $CPT$ invariance and assuming, for the sake of
simplicity, that only two terms dominate in a given decay amplitude,
the above asymmetry can be expressed as
\begin{equation}
a_{CP} =
\frac{2|A_1||A_2|\;\sin(\phi_1^W-\phi_2^W)\;\sin(\phi_1^S-\phi_2^S)}
{|A_1|^2 + |A_2|^2 +
2|A_1||A_2|\;\cos(\phi_1^W-\phi_2^W)\;\cos(\phi_1^S-\phi_2^S)}.
\end{equation}
It is clear that in order to produce large $a_{CP}$ the following
conditions will have to be satisfied: $(i)\;|A_1|\approx |A_2|$,
$(ii)\; \sin(\phi_1^W-\phi_2^W) \approx 1$ and $(iii)\;
\sin(\phi_1^S-\phi_2^S) \approx 1$.

\section{The standard model prediction}

In the SM, the decay $\bar{b} \rightarrow \bar{s} d\bar{d}$ receives
contributions only from penguin operators. Using the unitarity of the
Cabibbo-Kobayashi-Maskawa (CKM) matrix, one can write the decay
amplitude as \cite{core}
\begin{equation}
\label{asm}
A^{\rm SM}(B^+ \rightarrow \pi^+ K^0) =
-A\lambda^2(1-\lambda^2/2)\left[1+\rho e^{i\theta} e^{i\gamma}\right]
|P_{tc}| e^{i\delta_{tc}},
\end{equation}
where $\lambda = 0.22$ is the Wolfenstein parameter; $A \equiv
|V_{cb}|/\lambda^2 = 0.81 \pm 0.06$; $\gamma \equiv - {\rm Arg}
(V^*_{ub}V_{ud}/V^*_{cb}V_{cd})$ is the CKM weak phase; $\theta$ and
$\delta_{tc}$ are $CP$-conserving strong phases; $P_{tc} \equiv P_t^S
- P_c^S + P_t^W - P_c^W$ (the difference between top- and
charm-mediated strong and electroweak penguins); and finally, $\rho$
depends on the dynamics of the up- and charm-penguins. For calculating
$P_{tc}$ we employ the factorization technique, which has been
suggested to be quite reliable \cite{lu}. One can express $|P_{tc}|
\approx G_F f(\bar{C}_i) {\cal{F}}/\sqrt{2}$, where $f(\bar{C}_i)
\approx 0.09$ is an analytic function of the Wilson coefficients, and
${\cal{F}}=(m^2_{B_d}-m^2_{\pi})f_K F_{B\pi}$ with $F_{B\pi} = 0.3$.
The NLO estimate of ${\cal{B}} (B^\pm \rightarrow \pi^\pm K) \equiv
0.5[{\cal{B}} (B^+ \rightarrow \pi^+ K^0) + {\cal{B}} (B^- \rightarrow
\pi^- \bar{K^0})]$ varies in the range $(1.0-1.8)\times 10^{-5}$ for
$\rho = 0$ \cite{flma}.

Using Eq.~(\ref{asm}), the $CP$ asymmetry in the $B^+\rightarrow \pi^+
K^0$ channel is given by (neglecting tiny phase space effects)
\begin{equation} 
a_{CP}^{\rm SM} = -2\rho \sin\theta
\sin\gamma/(1+\rho^2+2\rho\cos\theta \cos\gamma).
\end{equation}
In the perturbative limit, $\rho = {\cal{O}}(\lambda^2 R_b) \sim
1.7\%$, where $R_b = |V_{ub}|/\lambda|V_{cb}| = 0.36 \pm 0.08$.  On
the other hand, rescattering effects \cite{core,rescatter}, such as,
$B^+ \rightarrow \pi^0 K^+ \rightarrow \pi^+ K^0$, {\em i.e.},
long-distance contributions to the up and charm penguins, can enhance
$\rho$ to as large as ${\cal{O}}(10\%)$ (this order-of-magnitude
estimate is based on Regge phenomenology).  As a result, $a_{CP}$ can
be as large as ${\cal{O}}(20\%)$. Therefore, if the upcoming
experiments measure a much larger $a_{CP}$, an undisputed evidence of
new physics with new phase(s) will be established.

\section{Effects of $R$-parity violation} 

\subsection{New diagram at tree level} 
In the minimal supersymmetric standard model, there are additional
penguins mediated by superparticles and they contain new phases. As a
result, $a_{CP}$ can go up to $\sim 30\%$ \cite{barbieri}. But
switching on $R$-parity-violating ($\Rsl$) $\lambda'_{ijk} L_i Q_j
D^c_k$ superpotential \cite{rpar,review} triggers new diagrams which
contribute to $B^+\rightarrow \pi^+ K^0$ (or other non-leptonic $B$
decays) at {\em tree level}. The interference between $\Rsl$ tree and
the SM penguins may generate large $a_{CP}$. Considering the current
upper bounds on the relevant $\lambda'$ couplings \cite{review}, it is
very much possible that the $\Rsl$ tree contributions are of the same
order of magnitude as the SM penguins.

\subsection{New weak phase}
A non-leptonic $B$ decay amplitude involves the product of the type
$\lambda'_{ij3}\lambda'^*_{ilm}$. The $\lambda'_{ijk}$ couplings are
in general complex. Even if a given $\lambda'$ is predicted to be real
in a given model, the phase rotations of the left- and right-handed
quark fields required to keep the mass terms real and to bring the CKM
matrix to its standard form automatically introduce a new weak phase
in this coupling, barring accidental cancellation. Thus the tree level
$R$-parity-violating $B^+ \rightarrow \pi^+ K^0$ amplitude in general
carries a new weak phase.

\subsection{New strong phase}
The isospin of a $|B^+\rangle$ state is 1/2, while that of a $|\pi^+
K^0\rangle$ state is either 1/2 or 3/2. The SM penguin operator does
not carry any isospin, while the $\Rsl$ tree operator carries an
isospin (0 or 1). As a consequence, the SM penguins produce $\pi^+
K^0$ final states always in the isospin 1/2 state, while the $\Rsl$
tree operator can produce the same final states in the isospin 3/2
state. The final state interaction between states with different
isospins may generate a relative strong phase between the SM penguins
and the $\Rsl$ tree diagrams.

\subsection{Computation of the new diagram}

To generate $B^+ \rightarrow \pi^+ K^0$ at tree level, consider that
$\lambda'_{i13}$ and $\lambda'_{i12}$ are the only non-vanishing
couplings. This constitutes a sneutrino ($\tilde{\nu}_i$) mediated
decay. The new amplitude can be written as (the negative sign in front
is just our convention)
\begin{equation}
\label{arsl}
A^{\Rslash}(B^+ \rightarrow \pi^+ K^0) = -
(|\lambda'_{i13}\lambda'^*_{i12}|/8 \tilde{m}^2){\cal{F}}
e^{i\gamma_{\Rslash}} \equiv -|\Lambda_{\Rslash}|
e^{i\gamma_{\Rslash}},
\end{equation}
where $\gamma_{\Rslash}$ denotes the weak phase associated with the
product of $\lambda'$s. The total amplitude then becomes
\begin{equation} 
A(B^+ \rightarrow \pi^+ K^0) = -A\lambda^2(1-\lambda^2/2) |P_{tc}|
e^{i\delta_{tc}} \left(1+\rho e^{i\gamma} e^{i\theta} + \rho_{\Rslash}
e^{i\gamma_{\Rslash}} e^{i\theta_{\Rslash}}\right),
\end{equation} 
where
\begin{equation}
\rho_{\Rslash} \equiv |\Lambda_{\Rslash}|/A\lambda^2(1-\lambda^2/2)
|P_{tc}|.
\end{equation}
It is important to note that $\rho_{\Rslash}$ is free from
uncertainties due to factorization. Numerically, $\rho_{\Rslash}$
could easily be order one.

\subsection{The $CP$ asymmetry} 
Assuming that $\rho_{\Rslash} \gg \rho$, one can write
\begin{equation}
\label{acp}
a_{CP} \approx -\frac{2\rho_{\Rslash} \sin\theta_{\Rslash}
\sin\gamma_{\Rslash}}{1 + \rho_{\Rslash}^2 + 2\rho_{\Rslash}
\cos\theta_{\Rslash} \cos\gamma_{\Rslash}}.
\end{equation}

To have a feeling of the size of $\rho_{\Rslash}$, we first choose
$\tilde{m} =$ 100 GeV throughout our analysis.  Employing the current
upper limits on $\lambda'_{i13}\lambda'_{i12}$ \cite{review}, we
obtain, for $i =$ 1, 2, and 3, $\rho_{\Rslash}~\ltap$ 0.17, 3.45, and
4.13, respectively. Therefore, it is possible to have a situation when
$\rho_{\Rslash} = 1$ (for $i =$ 2, 3). This implies that a 100\% $CP$
asymmetry is very much attainable, once we set $\gamma_{\Rslash} =
\theta_{\Rslash} = \pi/2$. We assert that such a drastic hike of $CP$
asymmetry constitutes a characteristic feature of $R$-parity violation
and it is hard to find such large effects in other places
\cite{chemtob}.

The minimum $\rho_{\Rslash}$ required to generate a given $a_{CP}$ is
given by (for $\rho = 0$)
\begin{equation}
\label{rhomin}
\rho_{\Rslash} ~\gtap ~(1-\sqrt{1-a_{CP}^2})/|a_{CP}|.
\end{equation}
Eq.~(\ref{rhomin}) has been obtained by minimizing $\rho_{\Rslash}$ with
respect to $\gamma_{\Rslash}$ and $\theta_{\Rslash}$ for a given
$a_{CP}$. Numerically,
\begin{equation}
\label{rhonum}
\rho_{\Rslash}~\gtap~ 1.0~(1.0), 0.50~(0.8), 0.33~(0.6), 0.21~(0.4),
0.10~(0.2);
\end{equation}
where the numbers within brackets refer to the corresponding
$a_{CP}$'s.

\subsection{New bounds on $\lambda'$ product couplings}

We should also be alert that $\rho_{\Rslash}$ does not become so large
that the prediction for ${\cal{B}} (B^\pm \rightarrow \pi^\pm K)$
exceeds the experimental upper limit. According to the latest CLEO
measurement, ${\cal{B}}^{\rm exp}(B^\pm \rightarrow \pi^\pm K) = (1.4
\pm 0.5 \pm 0.2)\times 10^{-5}$, which means ${\cal{B}}^{\rm exp} \leq
1.9\times 10^{-5}$ (1$\sigma$) \cite{cleo}. On the other hand,
${\cal{B}}^{\rm SM} \sim (1.0 - 1.8)\times 10^{-5}$ in view of the
present uncertainty of the SM (for $\rho = 0$; varying $\rho$ within
$0-0.1$ cannot change ${\cal{B}}^{\rm SM}$ significantly). Hence one
can accommodate a multiplicative new physics effect by at most a
factor 1.9 (at 1$\sigma$). One can check from the denominator of
Eq.~(\ref{acp}) that $\Rsl$ effects modify the SM prediction of the
branching ratio by a multiplicative factor $(1 + \rho^2_{\Rslash} +
2\rho_{\Rslash} \cos\theta_{\Rslash} \cos\gamma_{\Rslash})$. To
evaluate the maximum allowed value of $\rho_{\Rslash}$, we set one of
the two angles appearing in this factor to zero and the other to $\pi$
({\em i.e.}, we arrange for a maximum possible destructive
interference in the branching ratio). This leads to the conservative
upper bound
\begin{equation} 
\label{rhoup} 
\rho_{\Rslash} ~\ltap ~2.4;~~{\rm which
~implies}~~|\lambda'_{i13}\lambda'^*_{i12}| ~\ltap ~5.7\times
10^{-3}~~(1\sigma).
\end{equation}
For $i = 2, 3$, these limits are already stronger than the existing
ones.  Moreover, the existing bounds from semi-leptonic processes
necessarily depend on the exchanged squark masses and have been
extracted by assuming a common mass of 100 GeV for them. Present
Tevatron data disfavour such a low mass squark. On the contrary, our
bounds from hadronic $B$ decays rely on a more realistic assumption
that the exchanged sneutrinos have a common mass of 100 GeV. We also
note that the choice of phases that leads to Eq.~(\ref{rhoup})
predicts vanishing $CP$ asymmetry. If, on the other hand, we are
interested in finding the upper limit on $\rho_{\Rslash}$, keeping
$a_{CP}$ maximized with respect to $\gamma_{\Rslash}$ and
$\theta_{\Rslash}$, each of the two angles should be $\pi/2$.  This
way the interference term vanishes, and we obtain a stronger limit
$\rho_{\Rslash} ~\ltap ~1.0$. This, in conjunction with
Eqs.~(\ref{rhonum}) and (\ref{rhoup}), defines a range of the $\Rsl$
couplings to be investigated in the upcoming $B$ factories.

\section{Extraction of the CKM parameter $\gamma$} 

$B \rightarrow \pi K$ decays are expected to provide useful
information on $\gamma$, the least known angle of the unitarity
triangle. In the SM, the three angles ($\alpha$, $\beta$ and $\gamma$)
measured independently should sum up to $\pi$. However, if a given
channel is contaminated by new physics, it might lead to a wrong
determination. The $\Rsl$ couplings $\lambda'_{i13}$ and
$\lambda'_{i12}$, the only non-vanishing ones under consideration, do
not affect the $B_d \rightarrow \pi^+\pi^-$ and $B_d \rightarrow \Psi
K_s$ channels, which are used for the direct measurements of $\alpha$
and $\beta$, respectively. Once $\alpha$ and $\beta$ are measured this
way, $\gamma$ can be indirectly determined via the relation $\gamma =
\pi - \alpha - \beta$.

We now consider a direct measurement of $\gamma$, using the observable
\cite{flma} $ R \equiv [{\cal{B}}(B_d \rightarrow \pi^-K^+) +
{\cal{B}}(\bar{B}_d \rightarrow \pi^+K^-)]/ [{\cal{B}}(B^+\rightarrow
\pi^+ K^0) + {\cal{B}}(B^-\rightarrow \pi^- \bar{K^0})].  $ The
present experimental range is $R = 1.0 \pm 0.46$
\cite{cleo}. Neglecting rescattering effects, as a first
approximation, $\sin^2\gamma ~\ltap ~R$ \cite{flma} in the SM. Within
errors, it may still be possible that $R$ settles to a value
significantly smaller than one, disfavouring values of $\gamma$ around
$90^\circ$. This will certainly be in conflict if, for example,
$\gamma \approx 90^\circ$ is preferred by indirect determination.

Now we see the effects of $R$-parity violation. The SM bound is
modified to ($\rho = 0$)
\begin{equation}
\label{gbound3}
\sin^2\gamma ~\ltap~ R~(1 + \rho^2_{\Rslash} + 2\rho_{\Rslash}
\cos\theta_{\Rslash} \cos\gamma_{\Rslash}).
\end{equation}
Here we assumed that the left-handed selectron is somewhat heavier
than the sneutrino (due to the D-term contribution), so that the
neutral $B$ decay (which appears in the numerator of $R$) is not
significantly affected.  From Eq.~(\ref{gbound3}), we see that the
bound on $\gamma$ either gets relaxed or further constrained depending
on the magnitude of $\rho_\Rsl$ and the signs of $\gamma_{\Rslash}$
and $\theta_{\Rslash}$. For $\rho_\Rsl \approx 1$ and
$\gamma_{\Rslash} = \theta_{\Rslash} \approx \pi/2$, it turns out that
$\sin^2\gamma ~\ltap~ 2R$, and hence there is no constraint on
$\gamma$ once $R \gtap$ 0.5. Thus the lesson is that if one observes
large CP asymmetry in the $B^+ \rightarrow \pi^+ K^0$ channel,
extraction of $\gamma$ becomes an even more nontrivial exercise.

\section{Conclusions} 

We identified one distinctive feature of $R$-parity violation that it
can enhance both the CP asymmetry and the branching ratio
simultaneously. As a result it is easier to capture these effects
experimentally.  Our study is a prototype analysis in a particular
channel, which can be easily extended to a multichannel analysis
combining all kinds of $B \rightarrow \pi \pi, \pi K, D K $ modes. In
case no large {\em CP} asymmetry is observed, or no disparity between
${\cal{B}}^{\rm SM}$ and ${\cal{B}}^{\rm exp}$ is established, one can
place improved constraints on many $R$-parity-violating
couplings. Upcoming $B$ factories may very well turn out to be a
storehouse of many surprises!

\section*{Acknowledgements}
First, I thank Amitava Datta for a very fruitful collaboration. I also
thank Prof. H.V. Klapdor-Kleingrothaus and his team for organizing yet
another successful meeting at Ringberg Castle. I take this opportunity
to thank all my friends and colleagues at MPI/Heidelberg for their
warm hospitality during my stay this summer. Finally I acknowledge an
additional financial support of the Max-Planck Society, Germany, which
enabled me to attend this meeting.

\noindent {\em Note added:} 
After this manuscript had been written up, an interesting article
studying the impact of various new physics models, including the
$R$-parity-violating ones, on observables related to $B \rightarrow
\pi K$ decays appeared in the archive \cite{gnk}



\begin{thebibliography}{99}

\bibitem{bd} Bhattacharyya G and  Datta A 1999 \PRL {\bf 83} 2300.

\bibitem{core} Buras A, Fleischer R and Mannel T 1998 \NP {\bf B 533}
3; Fleischer R 1999 {\em Eur. Phys. J.} {\bf C6} 451; Fleischer R 1998
\PL {\bf B435} 221.

\bibitem{lu} Ali A, Kramer G and  L\"{u} C.-D 1998 \PR {\bf D58}
094009.

\bibitem{flma} Fleischer R and Mannel T 1998 \PR {\bf D57} 2752. 

\bibitem{rescatter} Falk A, Kagan A, Nir Y and Petrov A 1998 \PR {\bf
D57} 4290; Gronau M and Rosner J 1998 \PR {\bf D 57} 6843; Atwood D
and Soni A 1998 \PR {\bf D58} 036005.
 
\bibitem{barbieri} Barbieri R and  Strumia A 1997 \NP {\bf B508} 3.

\bibitem{rpar} Farrar G and Fayet P 1978 \PL {\bf B76} 575; 
Weinberg S 1982 \PR {\bf D26} 287;
Sakai N and Yanagida T 1982 \NP {\bf B197} 533;
Aulakh C and Mohapatra R 1982 \PL {\bf 119B} 136.

\bibitem{review} For recent reviews, see Bhattacharyya G {\em
hep-ph/9709395} and 1997 {\em Nucl. Phys. Proc. Suppl.}  {\bf 52A} 83;
Dreiner H {\em hep-ph/9707435}; Barbier R {\em et al.} {\em
hep-ph/9810232}.

\bibitem{chemtob} Chemtob M and Moreau G 1999 \PR {\bf D59} 116012.

\bibitem{cleo} CLEO Collaboration, Artuso M {\em et al.}, in {\em
Proceedings of the International Conference on High Energy Physics:
ICHEP'98, Vancouver, 1998} (CLEO CONF 98-20).

\bibitem{gnk} Grossman Y, Neubert M and Kagan A {\em hep-ph/9909297}.

\end{thebibliography}
\end{document}